\documentstyle[preprint,aps]{revtex}
                    \tighten
                    \begin{document}
                    \draft
                    \title{Models of Quantum Turing Machines} 
                    \author{Paul Benioff\\
                     Physics Division, Argonne National Laboratory \\
                     Argonne, IL 60439 \\
                     e-mail: pbenioff@anl.gov}
                     \date{\today}

                    \maketitle
                    \begin{abstract}  
          Quantum Turing machines are discussed and reviewed in this paper.
          Most of the paper is concerned with processes defined by a step
          operator $T$ that is used to construct a Hamiltonian $H$
          according to Feynman's prescription. Differences between these
          models and the models of Deutsch are discussed and reviewed.  It
          is emphasized that the models with $H$ constructed from $T$
          include fully quantum mechanical processes that take computation
          basis states into linear superpositions of these states.  

          The requirement that $T$ be distinct path generating is reviewed. 
          The advantage of this requirement is that Schr\"{o}dinger
          evolution under $H$ is one dimensional along distinct finite or
          infinite paths of nonoverlapping states in some basis $B_{T}$. 
          It is emphasized that $B_{T}$ can be arbitrarily complex with
          extreme entanglements between states of component systems.  

          The new aspect of quantum Turing machines introduced here is the
          emphasis on the structure of graphs obtained when the states in
          the $B_{T}$ paths are expanded as linear superpositions of states
          in a reference basis such as the computation basis $B_{C}$. 
          Examples are discussed that illustrate the main points of the
          paper. For one example the graph structures of the paths in
          $B_{T}$ expanded as states in $B_{C}$ include finite stage binary
          trees and concatenated finite stage binary trees with or without
          terminal infinite binary trees.  Other examples are discussed in
          which the graph structures correspond to interferometers and
          iterations of interferometers.
                    \end{abstract}

          \section{Introduction}
          Quantum computation is a field with much activity in recent
          years. Most work on models of quantum computers has been
          concerned with networks of quantum gates
          \cite{Deutsch89,EkJo,DiV}. Models based on quantum Turing
          machines \cite{Benioff82,Deutsch85,Benioff86,BeVa,BenioffQBE}
          have also been described.  Networks of quantum gates are
          appealing because they are similar to integrated circuits in that
          they are built up from simple elementary quantum gates.  Also
          working physical models of quantum computers are likely to have
          this network structure.  

          Networks of quantum gates are characterized by the fact that the
          complexity of the computation appears explicitly in the
          complexity of the layout and interconnections of the quantum
          gates and qubit lines making up the network.   Because of this
          networks of more than a few gates are hard to conceptualize as
          physical models. For complex networks heirarchical descriptions
          in terms of simpler component networks are useful.   Examples are
          the networks for the discrete Fourier transform \cite{EkJo} and
          Shor's Algorithm \cite{Shor,MigPer}.  

          Quantum Turing machines (QTMs) are quite different in that the
          complexity of the computation is not reflected in the physical
          model. Instead the complexity is reflected in the dynamical
          description of the model either as a unitary operator describing
          the possible finite time interval steps or in the description of
          the Hamiltonian for the model.  The physical model, consisting of
          a finite state head interacting with one (or if needed for
          convenience more than one) qubit lattice, remains fixed  for the
          different models. 

          In this paper the discussion will be limited to quantum Turing
          machines. An interesting aspect of these systems is that they can
          be expanded in different directions to illustrate different
          properties.  For example the description of the Hamiltonian can
          be easily expanded to include the presence of finite potential
          barriers on computation paths where the potential distribution is
          computable.  These result in reflections of the computation back
          along the computation paths \cite{Landauer}. An example of this
          has been discussed in detail elsewhere
          \cite{BenioffPRL,BenioffPRB}. Also the physical model of QTMs is
          a natural basis for expansion into a model of quantum robots
          interacting with an environment \cite{quant-ph9706012}.  

          In this paper step operators $T$ for each QTM are first defined
          by conditions on matrix elements in the computation basis. This
          follows the definition of Deutsch \cite{Deutsch85}. These
          operators can either be used to represent model steps occurring
          in a finite time interval, as was done by Deutsch, or they can be
          used to directly construct a Hamiltonian according to Feynman's
          prescription \cite{Feynman}.  This was the method used by
          Benioff. Differences between these two types of models are
          discussed in Section \ref{StepOp}.   

          Another definition of $T$ as a finite sum of elementary step
          operators is given in Section \ref{TSESO}.  This definition, much
          used by the author in earlier work \cite{Benioff86,BenioffQBE},
          has the advantage that description of a step operator with a
          specified set of elementary steps is easier under this definition
          than with the matrix element definition.  It is seen that, for
          the particular form of the unitary operators appearing in the
          elementary step sum, the matrix  element definition is more
          general in that final state correlations between changes in the
          head state, head position, and scanned qubit state are included. 
          Both definitions include initial state correlations and include
          steps that take computation basis states into linear
          superpositions of these states.

          In most of this paper QTM step operators $T$ will be used to
          construct model Hamiltonians according to Feynman's prescription
          \cite{Feynman}.  In this case the requirement that $T$ be
          distinct path generating much simplifies the dynamics of the
          system.  The main consequence of this requirement, which is 
          reviewed and discussed in Section \ref{DPG}, is that
          Schr\"{o}dinger evolution of the computation corresponds to one
          dimensional motion along paths of states in some basis $B_{T}$. 
          The paths are defined by iterations of $T$ and $T^{\dag}$ on
          states in $B_{T}$.  The dependence on $T$ is indicated by the
          subscript.  Basis dependent (matrix) and basis independent
          (operator) descriptions of this property are reviewed.  The form
          of the eigenfunctions and spectrum for Hamiltonians constructed
          from $T$s that are distinct path generating is also summarized. 
          A sum over paths representation of $e^{-iHt}$ for $H$
          constructed from step operators that are distinct path generating is
          described.  Section \ref{DPG} concludes with a review of the fact 
	  that there is no effective way to
 determine in general if a step operator is distinct path generating.

          Fully quantum mechanical computations are included because
          $B_{T}$ can be arbitrarily complex with extreme entanglements
          between the component system states.  Examples will be given to
          show this and to show that $T$ takes computation basis states
          into linear superpositions of computation basis states.  Thus
          distinct path generation is not limited to descriptions of a
          "classical Turing machine made of quantum components"
          \cite{EkJo}. Demonstration of this point, which was also made in
          \cite{BenioffQBE}, is one of the goals of this paper.

          The main novel feature introduced in this paper is an emphasis on
          the graph structure aspects of the paths in $B_{T}$ when the
          states in $B_{T}$ are expanded into superpositions of states in a
          fixed basis.  Stated differently the main interest here is not in
          the outcome of a process such as a quantum computation.  Instead
          the interest is in the graph properties of the computation paths
          relative to a fixed basis, such as the computation basis. Because
          of this aspect there is some overlap between this work and other
          work in the literature \cite{ExnerJPA,ExnerPRL,GratusetalJPA}  
          describing motion of systems in graphs of interconnected quantum
          wires.

          These properties are illustrated in Section \ref{Examples} by 
	  discussion of examples chosen to illustrate various points.  All
          the examples are fully quantum mechanical in that they take
          computation basis states into linear superpositions of these
          states.  The first example, the erasure? operator, is given to
          show that there are step operators, which at first glance appear
          to describe irreversible processes, in fact describe reversible
          processes.  The second example, product qubit transformations
          followed by add 1, describes generation of a linear
          superpositions of states corresponding to all numbers up to
          $2^{n-1}$ and adding 1 to each.  Depending on the initial states
          graph structures corresponding to one or more iterated binary
          trees with finite numbers of stages are generated.  Also a
          terminal infinite binary tree may or may not be appended.

          The third group of examples is introduced to show that closed
          graphs with loops are included.  In particular the graph
          structure of the examples corresponds to that of interferometers
          in coordinate space. The examples show that depending on the
          initial state, iterated loops corresponding to successive opening
          and closing of interferometer type structures are included.  It
          is seen that activities in each arm can be different although
          they must be such that differences in states in each arm must be  
	  removed coherently when the arms are closed.

          \section{The Physical Model}
          \label{TPM}
          The physical model used here corresponds to one-tape quantum
          Turing machines. Extension to machines with more than one tape,
          as in Bennett's description of reversible machines \cite{Bennett}
          is straightforward.  The model consists of a two-way infinite one
          dimensional lattice of systems each of which can assume states in
          a finite dimensional Hilbert space. if the space 
          is two dimensional, the systems are referred to as
          qubits.  This term will be used here even if the dimensionality
          is greater than two.  It is often convenient but not necessary to
          consider the lattice as spin systems, e.g. spin 1/2 systems for
          binary qubits.

          A head which can assume any one of a finite number of orthogonal
          states $\vert l\rangle$ with $l=1,2,\cdots ,L$ moves along the
          lattice interacting with qubits at or next to its location on the
          lattice.  Elementary QTM actions include one or more of (1) head
          motion one lattice site to the right or left, (2) change of the
          state of the qubit scanned or read by the head, (3) change of the
          head state.  What happens depends on the states of the head and
          scanned qubit. 

          Here the system states are all assumed to lie in a separable
          Hilbert space $\cal H$.  Based on the above description a
          particular basis, the computation basis, defined by the set of
          states $\{\vert l,j,\underline{s} \rangle\}$ and which spans $\cal
          H$, is used.  Here $l,j$ refer to the internal state and lattice
          position of the head.  The qubit lattice computation basis state
          $\vert \underline{s}\rangle =\otimes_{m=-\infty}^{\infty} \vert
          \underline{s}_{m}\rangle$ where $\underline{s}_{m} = 0\; \mbox{or}
          \; 1$ is the state of the qubit at lattice site $m$. From now on the
          computation basis will be denoted by $B_{C}$.

          In order to keep $\cal H$ separable (a denumerable basis), it is
          necessary to impose some condition on
          $\vert \underline{s}\rangle$.  Here it will be required that 
          $\underline{s}_{m} \neq 0$ for at most a finite number of values
          of $m$.  This condition, the $0$ tail state condition, is one of
          many that can be imposed to keep the basis denumerable.  Models
          of QTMs without tail conditions would have to be described by
          quantum field theory on a lattice to take account of the presence
          of an infinite number of degrees of freedom. 

          \section{The Step Operator}
          \label{StepOp}
          Models of QTMs in the literature
          \cite{Deutsch85,Benioff86,BenioffQBE,BenioffPRL} can be described
          by step operators $T$ that correspond 
	  to single steps of a
          computation.  In terms of matrix elements between states in
$B_{C}$, $T$ must satisfy a locality condition given by 
          superpositions of states in $B_{C}$:
          \begin{equation}
          \langle l^{\prime},j^{\prime},\underline{s^{\prime}}\vert T\vert
          l,j,\underline{s}\rangle =\langle \underline{s^{\prime}}_{\neq
          j}\vert \underline{s}_{\neq j}\rangle \langle
          l^{\prime},j^{\prime},\underline{s^{\prime}}_{j^{\prime}}\vert
          \tilde{T}\vert l,j,\underline{s}_{j}\rangle \label{Tdef}
          \end{equation}
	  Here $\vert \underline{s}\rangle =\vert \underline{s}_{\neq
j}\rangle \vert \underline{s}_{j}\rangle$ where $\vert
\underline{s}_{j}\rangle$ is the state of the site $j$ qubit.

          This condition states that single
          step changes in the state of the lattice qubits are limited to
          the qubit at the position of the head.

          The operator $\tilde{T}$ describes the interactions of the head
          with qubits on the lattice.  Since the interaction is localized
          to the qubit at the head position, $\tilde{T}$ can be described
          as an operator on the Hilbert space spanned by states of the form
          $\vert l,j,s\rangle$.  This corresponds to the head in state
          $\vert l\rangle$ at position $j$ and the site $j$ qubit in state
          $\vert s\rangle$.\footnote{In general the states have the form
          $\vert l,j,s,k\rangle$ that describe the head at site $j$ and the
          qubit at site $k$.  However since $T$ (and $\tilde{T}$) have
          nonzero matrix elements only for states of the form $\vert
          l,j,s,j\rangle$ where $k=j$, the two $j$ positions are combined.} 

          Another condition on $T$ is that head motion on the lattice is
          limited to at most one site in either direction.  Also all
          lattice sites are equivalent (space is homogenous). These
          conditions are given respectively by 
          \begin{eqnarray}
          \tilde{T} & = &  \sum_{j=-\infty}^{\infty}\sum_{\Delta =-1}^{1}P_{j+\Delta}\tilde{T}P_{j}
          \nonumber \\
          \langle l^{\prime},j^{\prime}+\Delta ,s^{\prime}\vert
          \tilde{T}\vert l,j^{\prime},s^{\prime}\rangle & = & \langle
          l^{\prime},j+\Delta ,s^{\prime}\vert \tilde{T}\vert l,j,s\rangle
          \label{Thomog}
          \end{eqnarray}
          for all $j,\; j^{\prime}$ and all $\Delta = -1,0,1$. $P_{j}$ is
          the projection operator for the head at site $j$.

          This definition differs in some inessential details from that
          given by  Deutsch \cite{Deutsch85} for QTMs.  One difference is
that the possibility that $\Delta =0$ is included. This addition is not
       essential in that it has been shown \cite{BeVa} that for
          every QTM that includes steps with no change in the head position
          there is an equivalent (slower) QTM which excludes these steps
          (i.e. $\Delta = -1,1$ only). Also systems with $\Delta =0$
          excluded are simpler to analyze.

          In the QTM models of Deutsch, $T$ describes the transformation
          associated with a finite time interval $t$. Thus $T$ is also
          required to be unitary.  Iterations of $T$ give the successive
          transformations at times that are integer multiples of $t$.  The
          implicit existence of Hamiltonians that generate these
          transformations was assumed.

          Models of QTMs described by Benioff
          \cite{Benioff86,BenioffQBE,BenioffPRL} differ from those of
          Deutsch in that $T$ is used directly to construct a Hamiltonian
          according to Feynman's prescription: \cite{Feynman}
          \begin{equation}
          H=K(2-T-T^{\dag}) \label{ham}
          \end{equation}
          where $K$ is a constant.  This definition has the property that
          if $T=Y$ where $Y$ describes free head motion with no interaction
          along the lattice ($Y\vert l,j,\underline{s}\rangle =\vert
          l,j+1,\underline{s}\rangle$), then $H$ is the kinetic energy of
          free head motion on the lattice.  As such it is equivalent to the
          symmetrized discrete version of the second derivative, $-
          (\hbar^{2}/2m)d^{2}/dx^{2}$. As defined $H$ is local as the head
          interacts only with qubits at or next to its location.  Also the
          requirement that $T$ be unitary is dropped.

          As noted above the main difference between the two types of
          models is that in those of Deutsch $T$ is unitary and it
          describes changes occurring in a finite time interval.  In the
          models of Benioff $T$ is used to directly construct a Hamiltonian
          $H$ with the changes in a finite time interval $t$ given by
          $e^{-iHt}$.  In this case $T$ need not be unitary.  

          One consequence of these differences is that in the models of
Benioff $e^{-iHt}$ is not
          local even though $H$ is local.  In the models of Deutsch
\cite{Deutsch85} $T$
represents the model evolution in a finite time interval and is supposed to
be local.  This requirement is not realistic in the sense that there is no
local Hamiltonian $H$ that satisfies $T=e^{-iHt}$ for finite $t$.   
This is easily seen by examination of the terms in a
          power series expansion of $e^{-iHt}$ where  $H$ has  a
          finite but nonzero spatial range.  

          Of course from a practical standpoint, even though $e^{-iHt}$ is
          not exactly local, it is approximately local in that matrix
          elements between sufficiently separated space positions are
          extremely small.  Otherwise it would not be possible to isolate
          systems for carrying out experiments. 

          \section{$T$ as a Sum of Elementary Step Operators} \label{TSESO}
	  
	  The conditions given on $T$ by Eqs. \ref{Tdef} and \ref{Thomog} for 
	  $T$ used in the
          Hamiltonian of Eq. \ref{ham} were done in a way to emphasize the
          similarities between the two types of QTM models.  However it is
          often quite useful to define $T$ as a finite sum
          \begin{equation}
          T=\sum_{l,s}T_{l,s} \label{Tgen}
          \end{equation}
          over elementary step operators $T_{l,s}$.  This method has been
          used by the author to model QTMs
          \cite{Benioff82,Benioff86,BenioffQBE,BenioffPRL} as this
          description makes it relatively clear what elementary steps
          are needed to accomplish a specific computation.  It also is
easier to see how the different elementary step operators interconnect in 
a QTM. 

          The elementary step operator $T_{l,s}$ describes the single step 
          that occurs in case the head is in state $\vert l\rangle$ and the
          qubit at the head location is in state $\vert s\rangle$.
          $T_{l,s}$ is defined by,
          \begin{equation}
          T_{l,s} = \gamma _{l,s}\sum_{j=-
          \infty}^{\infty}w_{l,s}Q_{l}v_{l,s}P_{s,j}u_{l,s}P_{j}
          \label{Tls}
          \end{equation}
          where $Q_{l},P_{j},P_{s,j}$ are respective projection operators
          for the head in internal state $\vert l\rangle$, at site $j$, and
          the site $j$ qubit in state $\vert s\rangle$.  $\gamma_{l,s}$ is
          a numerical constant.

          The operators $w_{l,s},v_{l,s},u_{l,s}$  act in the finite
dimensional Hilbert space of head states, the two dimensional qubit Hilbert
          space, and in the space of head lattice position states
          respectively where they  describe head and qubit state changes,
          and head motion.  The operators $w_{l,s},v_{l,s},u_{l,s}$ are
          unitary in their respective spaces, and $u_{l,s}$ satisfies the
          condition that $\langle j^{\prime}\vert u_{l,s}\vert j\rangle
          \neq 0$ only if $|j^{\prime}-j|=0,1$.   The possibility that
          these operators can be different for different values of $l,s$ is
          indicated by the subscripts.  

          The form of the equation for $T_{l,s}$ shows projection operators
          on computation basis states followed by unitary operators. 
          Depending on the process to be modelled it is sometimes useful to
          invert the order of the projection and unitary operators for one
          or more of the system components.  An example would be to replace
          $w_{l,s}Q_{l}$ by $Q_{l}w_{l,s}$ in Eq. \ref{Tls}.  This is the
          case if some component computation basis state is the desired
          final state.  This will be seen in the first and third examples
          studied.

          In most work using this form of $T,\: \gamma _{l,s} =1$ for all
          $l,s$ in the sum.  The case $0\leq \gamma_{l,s} \leq 1$ was
          considered elsewhere \cite{BenioffPRL}.  A detailed analysis of a
          simple case was done for which $\gamma_{l,s}<1$ corresponded to
          the introduction of potential barriers in the computation paths  
          \cite{BenioffPRB,BenioffPhysd}.  For this example the
          distribution and widths of the potential barriers in the paths
          was quasiperiodic \cite{DiVStn} and corresponded to a generalized
          substitution sequence \cite{BoGhJPA,KoNo,BuKe}.
           
          This definition includes $T_{l,s}$ that take computation basis
          states into linear superpositions of the basis states.  In
          particular $T\vert l,j,\underline{s}\rangle=\vert
          \underline{s}_{\neq j}\rangle
          \delta_{s,\underline{s}_{j}}T_{l,s}\vert l,j,s\rangle$ where
          \begin{equation}
          T_{l,s}\vert l,j,s\rangle
          =\gamma_{l,s}\sum_{l^{\prime},j^{\prime},s^{\prime}}\vert
          l^{\prime},j^{\prime},s^{\prime}\rangle \langle l^{\prime}\vert
          w_{l,s}\vert l\rangle \langle s^{\prime}\vert v_{l,s}\vert
          s\rangle \langle j^{\prime}\vert u_{l,s}\vert j\rangle
          \label{linsum}
          \end{equation}
          Here $w_{l,s}$ and $v_{l,s}$ can be any $L$ dimensional and $2$
          dimensional unitary operators respectively and $u_{l,s}$ can be
          such that the matrix elements $\langle j+\Delta\vert u_{l,s}\vert
          j\rangle$ are non zero for all three values of $\Delta$.

          The definition of $T$ as a step operator that satisfies Eqs.
          \ref{Tdef} and \ref{Thomog}  is somewhat more general than $T$
          defined as an $l,s$ sum by Eq. \ref{Tgen} with $T_{l,s}$ defined
          by Eq. \ref{Tls}. To see this one notes that any $T$ defined by
          Eqs. \ref{Tgen} and \ref{Tls} satisfies Eqs. \ref{Tdef} and
          \ref{Thomog}. 

          To examine the converse  let $T$ be a step operator that
          satisfies Eqs. \ref{Tdef} and \ref{Thomog}.  From these equations
          it follows that the corresponding operator $\tilde{T}$ and, as a
          result $T$, can always be expressed as an $l,s$ sum in the form  
          \begin{equation}
          T=\sum_{l,s}TP_{l,s} \label{Tls1}
          \end{equation}
          where 
          \begin{equation}
          P_{l,s}=\sum_{j=-\infty}^{\infty}Q_{l}P_{s,j}P_{j}
          \end{equation}
          where the projection operators are as defined in Eq. \ref{Tls}. 
          From this one sees that $TP_{l,s} =T_{l,s}$ if and only if
          \begin{equation}
          TP_{l,s} =\gamma_{l,s}w_{l,s}\otimes v_{l,s}\otimes
          u_{l,s}P_{l,s}. \label{TPls}
          \end{equation}
          This shows that the definition of $T_{l,s}$ given by Eq.
          \ref{Tls} is restricted to uncorrelated unitary transformations
          of the head state and position and qubit state.  That is the
          action of $T$ defined by Eqs. \ref{Tgen} and \ref{Tls} converts a
          computation basis state $\vert l,j,\underline{s}\rangle$ into
          $w_{l,s}\vert l\rangle u_{l,s}\vert j\rangle \vert 
	  \underline{s}_{\neq j}\rangle v_{l,s}\vert
          \underline{s}_{j}\rangle$ where $s=\underline{s}_{j}$. 
          Correlation with the
          initial state is included through the $l,s$ dependence of the
          different unitary operators in Eq. \ref{Tls}.

          \section{Distinct Path Generation} 
          \label{DPG}
          Here the property of distinct path generation for step operators
          will be reviewed and discussed.  This property is quite useful
          because the dynamics of any such process with a Hamiltonian given
          by Eq. \ref{ham} corresponds to one dimensional motion along
          distinct paths of states in some basis $B$.  In this case the
          complexity of the process is contained in the complexity of the
          states in $B$ and their ordering in paths. In particular the
          states in $B$ can describe arbitrary entanglements among states
          of component systems in the model.

          In general a path of states is an ordered set of states in a
          Hilbert space $\cal H$.  Here the interest is in paths generated
          by operators $T$ on $\cal H$ that represent the single steps of a
          process.  If $\psi_{1}$ is a state in $\cal H$, then the set of
          states $\{T^{n}\psi_{1},\: (T^{\dag})^{-m}\psi_{1}\}$ for
          integral $m,n$ where $n\geq 0,\: m<0$ represents a path of
          states. The ordering of the states in a path is conveniently
          expressed by the notation $\vert k,i\rangle \sim T^{k}\psi_{1}$
          for integral $k\geq 0$ and $\vert k,i\rangle 
	  \sim (T^{\dag})^{-k}\psi_{1}$
          for $k<0$.  The relation $\sim$ means equality up to
          normalization. The path label is denoted by $i$ where $\vert
        0,i\rangle =\psi_{1}$ and $k$ denotes the path position of the state. 

          The state $T^{n}\psi_{1}$ for $n\geq 0$ is a terminal state if
          $T^{n+1}\psi_{1}=0$.  Similarly $(T^{\dag})^{-m}\psi_{1}$ for
          $m<0$ is terminal if $(T^{\dag})^{-m+1}\psi_{1}=0$.  Paths are
          two way infinite or cyclic if they have no terminal states, one
          way infinite if they have one terminal state, and finite with
          distinct ends if they have two terminal states.

          The goal of the above definitions of paths is to satisfy the
          condition that iterations of $T$ or $T^{\dag}$ on states describe 
          successive steps of a quantum computation in the forward or
          backward direction respectively. So far this requirement is not
          satisfied because any state $\vert k,i\rangle$ in a path may have
          nonzero overlap with other states $\vert j,i\rangle$ in the path. 
          Thus state $\vert k,i\rangle$ which is supposed to correspond to
          completion of the $kth$ step may also correspond to completion of
          the $jth$ and other steps.

          To remove this possibility it is required that all the states in
          any path must be pairwise orthogonal.  If this condition is
          satisfied then the state $\vert k,i\rangle$ represents
          unambiguously the completion of the $kth$ step in the forward
          direction for $k\geq 0$ and in the backward direction for $k<0$.  
          Another condition that must be satisfied is that for each $n\geq
          0$ and $m<0$, $T^{\dag}T^{n+1}\psi_{1}=T^{n}\psi_{1}$ and
          $T(T^{\dag})^{-m}\psi_{1}=(T^{\dag})^{-m-1}\psi_{1}$.  This
          condition expresses the requirement that motion one step
          backwards from the state corresponding to completion of the
          $n+1st$ step should give the state corresponding to completion of
          the $nth$ step. Also motion one step forward from the $-mth$ step
          in the backward direction should give the state corresponding to
          the $-m-1st$ backward step (recall that $m<0$).

          A third condition is that if $\psi_{2}$ is a state that is
          orthogonal to all states in the path defined above for $\psi_{1}$, 
	  denoted from now on as path $1$, then all states in path $2$,
generated as defined above with $\psi_{2}$ replacing $\psi_{1}$, should be
          orthogonal to all states in path $1$.  By convention $\psi_{2}$
          is in no path if $T\psi_{2} = T^{\dag}\psi_{2} = 0$.  This
          expresses the intuitive requirement that computations started out
          on initial states that are orthogonal must maintain the
          orthogonality through all steps, forward and backward, of the
          computation.  Otherwise one could not associate an outcome with a
          specific input.

          Mathematically these conditions are expressed by requiring that
          $T$ be distinct path generating.  That is there must exist some
          basis $B$ such that for each $\vert b\rangle$ in $B$ if $T\vert
          b\rangle \neq 0$ then, up to normalization, $T\vert b\rangle$ is
          a state in $B$.  Similarly if $T^{\dag}\vert b\neq 0$, then, up
          to normalization, $T^{\dag}\vert b\rangle$ is in $B$.  It follows
          from this condition that all states in a path are pairwise
          orthogonal or both $T$ and $T^{\dag}$ are the identity on some
          $\vert b\rangle$.  Up to normalization means that the normalized
          state $T\vert b\rangle /|\langle b\vert T^{\dag}T\vert
          b\rangle|^{1/2}$ is in $B$.  A similar condition holds for
          $T^{\dag}$.

          The above condition of pairwise orthogonality of states within a
          path combined with the condition of orthogonality of all states
          in one path with those in another path can be expressed by
          $\langle j,i\vert k,i^{\prime}\rangle =0$ unless $j=k$ and
          $i=i^{\prime}$.  Here $\vert k,i\rangle$ and$\vert
          j,i^{\prime}\rangle$ are states in $B$.  The condition that
 states generated by iteration of $T$ and $T^{\dag}$ remain within a path
          is expressed by $T\vert k,i\rangle \neq 0 \rightarrow T\vert
          k,i\rangle \sim \vert k+1,i\rangle$ and $T^{\dag}\vert k,i\rangle
          \neq 0\rightarrow T^{\dag}\vert k,i\rangle \sim \vert k-
          1,i\rangle$. 

          These conditions for distinct path generation can be combined
          into one condition.  This is that $T$ must be such that there
          exists a basis $B$ for which each row and each column of the
          matrix of $T$ in $B$ contains at most one nonzero element.  If
          this condition is satisfied, it is always possible to choose $B$
          so that the matrix elements of $T$ are positive real numbers.  
          In addition the matrix elements are bounded from above, since $T$
          is bounded, and from below, for technical reasons (noted in the
          next section).

          It is convenient to let $B_{T}$ denote any basis for which $T$
          (and $T^{\dag}$) is distinct path generating.  In general if $T$
          is distinct path generating, there are a great many bases for
          which $T$ is distinct path generating.  This may be the case if
          there is more than one copy of paths of the same type (two way
          infinite, finite of the same length, etc.,) present.  For the
        models of quantum computers considered here there are either no copies
          or infinitely many copies of the same type present. This is a
          consequence of the spatial homogeneity of $T$. 

          \subsection{Basis Independent Description of Distinct Path
          Generation}

          The above description of distinct path generation is basis
          dependent as it is given in terms of matrix elements of $T$ in
          some basis.  An equivalent basis independent operator theoretic
          description is also possible.  It has been shown elsewhere
          \cite{BenioffQBE,BenioffPRL} that the conditions given above for
          distinct path generation are equivalent to the requirement that
          $T$ is a direct sum of weighted shifts \cite{Halmos}.  That is
          \begin{equation}
          T=UD=\oplus_{i}U_{i} D_{i} \label{Tdirsum}
          \end{equation}
          where for each $i$ $U_{i}$ is either a bilateral shift, a
          unilateral shift, the complement of a unilateral shift, a finite
          shift or a cyclic shift. Many copies of each type can be present
          and some types may be absent.  The types of paths associated with
          each of these shift types are shown in Figure 1.  Note that the
          only shifts that are unitary are the bilateral and cyclic shifts.
            
          This follows from the decomposition theorem for power partial
          isometries \cite{HW} since $U$ is a power partial isometry and
          $D$ is a diagonal operator \cite{Halmos}.  The $i$ sum
          corresponds to a sum over distinct path subspaces ${\cal H}_{i}$
          where ${\cal H}_{i}$ is spanned by the basis states in the path
          generated by $U_{i}$ and its adjoint.  That is if $\vert
          j,i\rangle$ is such a state in ${\cal H}_{i}$ then either $\vert
          j,i\rangle$ is terminal for $U$ or $U\vert j,i\rangle =U_{i}\vert
          j,i\rangle =\vert j+1,i\rangle$. A similar condition holds for
          $U^{\dag}$.  Here the set of states $\{\vert j,i\rangle\}$ (plus
          any additional basis states that span the null space for $T$, if
          such a space exists) form a distinct path generating basis for
          $T$.

          It should be noted that the converse implication does not hold. 
          That is if $U$ is a power partial isometry, it does not follow
          that $U$ is distinct path generating in some basis.  The reason
          is that the unitary component of the decomposition may not be
          distinct path generating in any basis \cite{BenioffQBE}. 
          Distinct path generation does follow if the unitary component of
          $U$ in the decomposition is either empty or is a bilateral or
          cyclic shift. 

          The operator $D$ is self adjoint and bounded with eigenstates
          $\vert j,i\rangle$.  That is $D\vert j,i\rangle =D_{i}\vert
          j,i\rangle =d_{j,i}\vert j,i\rangle$ where $d_{j,i}$ is a
          positive real number.  For technical reasons it is useful to
          require that $d_{j,i} > \epsilon >0$ for all $j,i$ \cite{Halmos}. 
          This avoids such functions as $d_{j,i}=1/j$ on infinite paths. 
          Note that the operator $D$ accounts for the possible loss of
          normalization referred to for $T$. Additional details on $U$ are
          given in \cite{HW,BenioffQBE}.

          \subsection{Eigenfunctions, Spectrum of H}
          \label{eig}
          If $T$ is distinct path generating the eigenfunctions and
          energy spectrum of the Hamiltonian of Eq. \ref{ham} all
          correspond to one dimensional motion on a path of states in
          $B_{T}$. If $T =\sum_{i}U_{i}$ with $D=1$ the motion is free
          except for the possible presence of infintely high potential
          walls located adjacent to terminal path states.  Details are given in
          \cite{BenioffQBE}.  If no walls are present (path a in Figure 1) 
	  the eigenfunctions
          are given by $\psi_{k} =\sum_{j=-\infty}^{\infty}e^{ikj}\vert
          j,i\rangle$ where $k$ is the momentum.  The $B_{T}$ path state
$\vert j,i\rangle$ is
          defined at the beginning of section \ref{DPG}. The energy
          eigenvalues are given by $E=2K(1-\cos k)$ where $k$ can take any
          value between $-\pi$ and $\pi$.

          If one wall is present the eigenstates describe standing waves
          reflected off the potential wall.  For reflection to the left
(path b in Figure 1) eigenstates are given by $\psi_{k}=\sum_{j=-\infty}^{b} 
\sin k(b-j)\vert j,i\rangle$ where $b$ is the wall location ($T\vert
          b-1,i\rangle =0$).  For reflection to the right (path c in Figure 1)
          $\psi_{k}=\sum_{j=a}^{\infty}\sin k(j-a)\vert j,i\rangle$ where
          $a$ is the wall location ($T^{\dag}\vert a+1,i\rangle =0$).  The
          eigenvalues are the same as for the free case above. 

          For finite paths (path d in Figure 1) the eigenstates describe bound 
state motion between two reflecting walls.  For the walls located at $a$ and
          $b$ with $a<b$ the eigenstates are given by
          $\psi_{k}=\sum_{j=0}^{b-a} \sin k(b-a-j)\vert a+j,i\rangle$. The
          energy is given by $E=2K(1-\cos k)$ where $k=2\pi m/(b-a)$ with
          $m=1,2,\cdots ,b-a-1$. A similar expression holds for cyclic paths.

          For the case in which $D\neq 1$ finite potentials of different
          heights and widths can be present in addition to the reflecting
walls at path terminal states.  The eigenstates and spectra of $H$ are 
much more complex
          in that reflections and transmissions occur at the path locations
          of the potentials \cite{Landauer}. A specific example analyzed
          elsewhere \cite{BenioffPRL,BenioffPRB,BenioffPhysd} showed a
          complex band structure for the spectrum of $H$.

          \subsection{Sum Over Paths Representation}

          It is of interest to examine a sum over paths representation of
          the unitary evolution operator $e^{-iHt}$ to show the relation
          between the paths defined by iteration of $T$ and $T^{\dag}$ and
          those in the path sum.  $T$ is assumed to be distinct path
          generating.  The representation is based on a straightforward
          power series expansion of $e^{-iHt}$. In the context of this
          paper, representation as a sum or integral over action weighted
          paths \cite{FeynmanH}  is not used as it appears to require more
          development.

          The time development of the state of a QTM under the action of a
          Hamiltonian $H$ can be expressed in any basis $B$ by
          \begin{equation}
          \Psi (t) = \sum_{b^{\prime},b\epsilon B} \vert b^{\prime}\rangle
          \langle b^{\prime}\vert e^{-iHt}\vert b\rangle \langle b\vert
          \Psi (0)\rangle.
          \end{equation}
          Here $\Psi (t)$ and $\Psi(0)$ are the system states at times $t$
          and $0$ and $\langle b^{\prime}\vert e^{-iHt}\vert b\rangle$ is
          the amplitude that a QTM in state $\vert b\rangle$ is in state
          $\vert b^{\prime}$ after a time interval $t$. 

          Use of the power series expansion gives
          \begin{equation}
          \langle b^{\prime}\vert e^{-iHt}\vert b\rangle
          =\sum_{n=0}^{\infty} \frac{(-it)^{n}}{n!}\langle b^{\prime}\vert
          H^{n}\vert b\rangle.
          \end{equation}
          Expansion of the matrix element by inserting a complete set of
          $B$ states between each $H$ factor gives
          \begin{equation}
          \langle b^{\prime}\vert H^{n}\vert b\rangle =
          \sum_{b_{1},b_{2},\cdots b_{n-1}}\langle b^{\prime}\vert H\vert
          b_{n-1}\rangle \cdots \langle b_{2}\vert H\vert b_{1}\rangle
          \langle b_{1}\vert H\vert b\rangle.
          \end{equation}
          This is equivalent to a sum over paths $p$ of states  in $B$ of
          length $n+1$ that begin with $\vert b\rangle$ and end with $\vert
          b^{\prime}\rangle$:
          \begin{equation}
          \langle b^{\prime}\vert H^{n}\vert b\rangle =\sum_{p}^{\prime}
          \langle p(n+1)\vert H\vert p(n)\rangle \cdots \langle p(3)\vert
          H\vert p(2)\rangle \langle p(2)\vert H\vert p(1)\rangle.
          \label{pathsum}
          \end{equation}
          The prime on the sum means the sum is restricted to length $n+1$
          paths with the initial and terminal restrictions noted above.

          The equations hold for arbitrary Hamiltonians  $H$ and bases $B$. 
          If $H$ is given by Eq. \ref{ham} and $T$ is distinct path
          generating in  $B$ (i.e. $B=B_{T}$) the matrix element $\langle
          b^{\prime}\vert H^{n}\vert b\rangle =0$ for all $n$ unless $\vert
          b^{\prime}\rangle$ and $\vert b\rangle$ are in the same $T$ path. 
          $T$ paths are the paths defined earlier by iteration of powers of
          $T$ and $T^{\dag}$ on states of $B_{T}$. 

          If $\vert b^{\prime}\rangle$ and $\vert b\rangle$ are in the same
          $T$ path, the above shows that  $\langle b^{\prime} \vert e^{-
          iHt}\vert b\rangle$ is equal to a sum of path amplitudes over
          paths of all lengths within the $T$ path containing $\vert
          b^{\prime}\rangle$ and $\vert b\rangle$. Each path in the sum
          describes 1-D motion within the $T$ path.  This includes motion in
          both directions with reflections from any terminal states or
          potentials encountered \cite{BenioffPRL}.  

          \subsection{Effective Determination of Distinct Path Generation}
          \label{EDDPG}

          The question can be asked if there is any effective way to
          determine if a step operator $T$ that satisfies Eqs. \ref{Tdef}
          and \ref{Thomog} is distinct path generating in some basis
          $B_{T}$.  That is, does there exist any algorithm or computation
          which can decide in a finite number of steps for any $T$ that
          satisfies Eqs. \ref{Tdef} and \ref{Thomog} whether $T$ is or is
          not distinct path generating?

          The first step is to note that if $T$ is distinct path generating
          in the computation basis $B_{C}$, then this can be decided
          effectively.  To see this note that if $T$ satisfies Eqs.
          \ref{Tdef} and \ref{Thomog}, then, because of space homogeneity,
          it is sufficient to search through the matrix elements of
          $\tilde{T}$ with $j$ set at a fixed value, say $j=0$. The search
          involves deciding if each row and column of the matrix for
          $\tilde{T}$ has at most one nonzero element. Since for fixed $j$
          the matrix is finite dimensional, this can be effectively decided
          provided there is an effective procedure for determining if the
          matrix elements of $\tilde{T}$ are zero or nonzero.

          The more general situation in which $T$ is distinct path
          generating in a basis $B_{T}$ different from $B_{C}$ was examined
          elsewhere \cite{BenioffQBE}. There it was shown using the
          operator theoretic description that in general no effective
          procedure exists. For the matrix description of $T$ one would
          need to examine the states $T^{n}\vert l,0,s\rangle$,
          $(T^{\dag})^{m}\vert l,0,s\rangle$ for all $m,n$ and all $l,s$ to
          see if the resulting states can be organized into a basis for
          which $T$ is distinct path generataing.  This procedure is not
          effective as there are an infinite number of values of $n,m$.  

          The fact that there is no effective proceedure for general $T$
          does not prevent one from studying many examples for which
          distinct path generation on bases different from $B_{C}$ can be
          demonstrated.  For these examples one method of proving that $T$
          is distinct path generating is to prove that $T$ is a power
          partial isometry or direct sum of shifts.  Another method that is
          sometimes useful is the direct construction of the states in
          $B_{T}$. Use of these methods will be seen in the following
          examples. 

          \section{Examples} \label{Examples}

          One goal of studying examples is to show the large diversity of
          QTMs with step operators $T$ that are distinct path generating
          for bases $B_{T}$ that are different from $B_{C}$.   All examples
          discussed here have this property.  The second example is chosen
          to emphasize this property. It also is an example where the
          activity or sequence of steps carried out on each computation
          path in the computation basis is different for the different
          paths.  It is a computation of the form $\vert
          \underline{0}\rangle \rightarrow \sum _{s=0}^{2^{n}-1}c_{s}\vert
          \underline{s}\rangle \rightarrow \sum _{s=0}^{2^{n}-1}c_{s}\vert
          \underline{f(s)}\rangle$.  Here $\underline{s}$ and
          $s=\sum_{l=1}^{n}\underline{s}(l)2^{l-1}$ denote a length $n$
          binary sequence and the corresponding number. The constant $0$
          sequence is denoted by $\underline{0}$.  The computed function
          $f$ is the simple one-one function "add $1 \bmod 2^{n}$" where
          $n$ is arbitrary.    

          Another goal is the study of the structure of the graphs obtained
          by expansion of the states in the $B_{T}$ paths as superpositions
          of states in a reference basis such as $B_{C}$.  Many of the
          graphs for the above example have the form of finite binary trees
          that may be iterated by appending other trees to the terminal
          branch lines from the preceeding tree.  These graphs have the
          property that they are open in that they contain no closed loops. 

          The third example (actually two examples) was chosen to
          illustrate a closed loop graph with one input and one output
          line.  The graph structure of paths of states in $B_{T}$, when
expanded in terms of states in $B_{C}$, is that
          of an interferometer in coordinate space.  For these examples the
          emphasis is on the graph structure and not on computation.  The
          first example is very simple with identical activity occurring in
          the two arms.  The second example is slightly more complex in
          that the activity is different in each of the two arms.  However,
          differences in path states resulting from these activity
          differences must be removed coherently before the interferometer is 
   closed.   This corresponds to bringing the two arms together in coordinate
          space interferometers.  

          The main purpose of the first example is to show that some step
          operators have different properties than they appear to have at
          first glance.  For this example the step operator appears to
          describe an irreversible process of erasure in a string of
          qubits.  However it describes a quite different reversible
          process for which $T$ is distinct path generating. 

          \subsection{The Erasure? Operator}
          \label{erasop}
          Let the step operator $T$ be defined by  
          \begin{equation}
          T=\sum_{j=-\infty}^{\infty}\frac{\sigma_{x}P_{1,j}
          +P_{0,j}}{\sqrt{2}}uP_{j}. \label{Teras}
          \end{equation}
          The Pauli operator $\sigma_{x}$ satisfies $\sigma_{x}P_{i,j}
          =P_{i \oplus 1,j}\sigma_{x}$ where $i\oplus 1$ denotes addition
          mod 2. The operator $u$ is a shift operator for the head on the
          lattice, $uP_{j}=P_{j+1}u$. No internal head states are needed.  

          Iteration of this operator on a computation basis state moves the
          head from left to right on the lattice converting each qubit
          state $\vert 1\rangle$ to $(1/\sqrt{2})\vert 0\rangle$ and
          leaving $\vert 0\rangle$ alone other than changing the
          normalization to $(1/\sqrt{2})\vert 0\rangle$. Thus iteration of
          $T$ appears at first glance to describe an irreversible erasure
          along the lattice of qubits changing all $1s$ to $0$ and leaving
          $0$ alone.

          Appearances are deceiving, though, because $T$ is in fact
          distinct path generating on a basis $B_{T}\neq B_{C}$.  This is
          the reason for the question mark in the subsection title. This
          can be seen by rewriting $T$ in the form 
          \begin{equation}
          T=\sum_{j=-\infty}^{\infty} \sqrt{2}P_{0,j}P_{+,j} uP_{j}.
          \label{eras?op}
          \end{equation} 
          This follows from the fact that $(1+\sigma_{x})/2$ is the
          projection operator $P_{+}$ on the state $\vert +\rangle
          =1/\sqrt{2}(\vert 0\rangle +\vert 1\rangle)$.

          It is straightforward to see that 
          $T$ and $T^{\dag}$ are power partial isometries
          \cite{HW,BenioffQBE}.  This follows from the fact that for each
          $n=1,2,\cdots$, $(T^{\dag})^{n}T^{n}$ and $T^{n}(T^{\dag})^{n}$ 
          given by
          \begin{eqnarray}
          (T^{\dag})^{n}T^{n} & = & \sum_{j=-
          \infty}^{\infty}\prod_{k=0}^{n-1}P_{0,j+k}P_{j} \nonumber \\
          T^{n}(T^{\dag})^{n} & = & \sum_{j=-\infty}^{\infty}
          \prod_{k=1}^{n}P_{+,j-k}P_{j}
          \end{eqnarray}
          are projection operators.  Also for each $m,n$ the operators
          $(T^{\dag})^{n}T^{n}$ and $T^{m}(T^{\dag})^{m}$ commute
          \cite{HW,BenioffQBE}.  

          The direct sum decomposition of $T$, Eq. \ref{Tdirsum} into
          shifts (as $D=1$) contains no unitary components. All paths are
          either finite or one way infinite from the left.   The operators
          $U_{i}$ are either a complement of a unilateral shift (a
          coisometry) or a finite shift.  The absence of unilateral shifts
          or bilateral shifts is a consequence of the $0$ state tail
          condition in the definition of $\cal H$.  The eigenstates and
          eigenvalues are as described in subsection \ref{eig}.

          For the one way infinite path the path states in $B_{T}$ have the
          form 
          \begin{equation}
          \vert n,i\rangle =\vert n\rangle \otimes \theta_{n}
          \label{stateras}
          \end{equation}
          Here $\vert n\rangle$ denotes the head at site $n$. The state of
          the lattice qubits $\theta_{n}$ is given by 
          \begin{equation}
          \theta_{n} = \vert \underline{0}_{\leq n}\rangle
          \otimes_{j=n+1}^{b-1}\vert +\rangle_{j}\otimes \vert -\rangle_{b}
          \otimes \alpha_{>b}.
          \end{equation}
          Here $\vert \underline{0}_{\leq n}\rangle$ denotes the constant
          $0$ sequence for all lattice sites $\leq n$ and $\vert
          \pm\rangle_{j}$ denotes the state $(1/\sqrt{2})(\vert 0\rangle
          \pm \vert -\rangle)$ for the site $j$ qubit.  The path state
          $\vert b,i\rangle$ is terminal for $T$. The value of $b$ is 
	  arbitrary.  The
          state $\alpha_{>b}$ denotes any state of the qubits at sites $>b$
          consistent with the $0$ state tail condition.  The arbitrariness
          in $\alpha_{>b}$ is possible because the head never enters the
          lattice region to the right of site $b$.  

          For finite paths between and including lattice sites $a$ and $b$,
          the path states in $B_{T}$ have the same form as in Eq.
          \ref{stateras} with $a\leq n\leq b$. 
          The state of the lattice qubits $\theta_{n}$ is given by 
          \begin{equation}
          \theta_{n} = \alpha_{\leq a-2}\otimes \vert 1\rangle_{a-1}
          \otimes \vert \underline{0}_{[a,n]}\rangle \otimes_{j=n+1}^{b-
          1}\vert +\rangle_{j}\otimes \vert -\rangle_{b} \otimes
          \alpha_{\geq b+1}.
          \end{equation}
          Here $\vert \underline{0}_{[a,n]}\rangle$ denotes the constant 0
          state for qubits on sites between $a$ and $n$, $\alpha_{\geq
          b+1}$ is as defined before, and $\alpha_{\leq a-2}$ is an
          arbitrary state of the qubits on lattice sites $\leq a-2$
          consistent with the $0$ state tail condition. The head never
          moves outside the lattice region $[a,b]$ as $T\vert b,i\rangle
          =T^{\dag}\vert a,i\rangle =0$.

          \subsection{General Product Qubit Transformation and Add 1}

          This example is given to show that step operators that are
          distinct path generating include computations in which linear
          superpositions of input states each corresponding to a different
          numerical input are generated.  In addition the computation
          activity, such as the number and actions of the elementary steps,
          depends on the numerical input.  

          The QTM considered here moves down a string of qubits each in
          state $\vert 0\rangle$ between two markers carrying out an
          arbitrary but fixed unitary transformation $v$ on each qubit to
          generate the lattice qubit state $\Psi =\otimes _{j}v\vert
          0_{j}\rangle$ for all the qubits between the markers.  Next the
          QTM adds $1 \bmod 2^{n}$ to each length $n$ state $\vert
          \underline{s}\rangle$ in the superposition $\Psi$.   For each
          $\vert \underline{s}\rangle$ the computation begins with the head
          moving toward the marker region and ends with the head moving
          away from the marker region with no further changes in the
          lattice.  The number $n$ is determined by
          the separation of the two markers.

          In order to accomodate marker states each qubit is assumed to be
          a ternary system with a basis $\vert i\rangle ,\: i=0,1,2$
          spanning the Hilbert space of each qubit.  In this case the two
          dimensional unitary operator $v$ is expanded to be the identity
          when acting on $\vert 2\rangle$ or $v\rightarrow P_{2}\oplus v$ where
          $P_{2}$ is the projection operator for the state $\vert
          2\rangle$.  In the following $v$ is assumed to be so expanded.

          A step operator $T$ that implements this model is the sum of 9
          terms.  It is given by 
          \begin{eqnarray}
          T  & = &  \sum_{j=-\infty}^{\infty}\left[ \begin{array}{c}
          Q_{0}P_{0,j}u \\ 1 \end{array} + \begin{array}{c} wQ_{0}P_{2,j}u
          \\ 2 \end{array} + \begin{array}{c} Q_{1}v_{j}P_{0,j}u \\ 3
          \end{array}  + \begin{array}{c} wQ_{1}P_{2,j}u^{\dag} \\ 4
          \end{array}  + \begin{array}{c} Q_{2}\sigma_{x,j}P_{1,j}
          u^{\dag} \\ 5 \end{array} \right. \nonumber \\
	  & & \mbox{} + \left. \begin{array}{c}
          wQ_{2}\sigma_{x,j}P_{0,j}u \\ 6 \end{array} +\begin{array}{c}
          wQ_{2}P_{2,j}u \\ 7 \end{array} + \begin{array}{c} Q_{3}P_{0,j}u
          \\ 8 \end{array}  + \begin{array}{c} wQ_{3}P_{2,j}u \\ 9
          \end{array} \right] P_{j}. \label{Tadd1}
          \end{eqnarray}
          In the above sum for $T$ $Q_{i}$ with $i=0,1,2,3$ is the
          projection operator for the head in state $\vert i\rangle$ and
          $w$ shifts the head state by $1 \bmod 4$.  That is
          $wQ_{i}=Q_{i+1}w \bmod 4$.  The projection operator $P_{j}$ for
          the head at site $j$ is at the end of the expression and $v_{j}$
          is the expanded unitary operator described above for the site $j$
          qubit.

          The numbers under each term denote the term number for each of
          the 9 terms.  They are present to facilitate discussion of the
          action of $T$ and $T^{\dag}$.  $T^{\dag}$ is given by
          \begin{eqnarray}
          T^{\dag}  & = &  \sum_{j=-\infty}^{\infty}\left[ \begin{array}{c}
          Q_{0}P_{0,j}P_{j}u^{\dag} \\ 1^{\dag} \end{array} +
          \begin{array}{c} Q_{0}w^{\dag}P_{2,j}P_{j}u^{\dag} \\ 2^{\dag}
          \end{array} + \begin{array}{c} Q_{1}P_{0,j}v_{j}^{\dag}P_{j}u^{\dag}
          \\ 3^{\dag} \end{array}  + \begin{array}{c}
          Q_{1}w^{\dag}P_{2,j}P_{j}u \\ 4^{\dag} \end{array} + 
	  \begin{array}{c} Q_{2}P_{1,j}\sigma_{x,j}P_{j} u
          \\ 5^{\dag} \end{array} \right. \nonumber \\
	  & & \mbox{} + \left. \begin{array}{c}
          Q_{2}w^{\dag}P_{0,j}\sigma_{x,j}P_{j}u^{\dag} \\ 6^{\dag}
          \end{array} +\begin{array}{c} Q_{2}w^{\dag}P_{2,j}P_{j}u^{\dag}
          \\ 7^{\dag} \end{array} + \begin{array}{c}
          Q_{3}P_{0,j}P_{j}u^{\dag} \\ 8^{\dag} \end{array}  +
          \begin{array}{c} Q_{3}w^{\dag}P_{2,j}P_{j}u^{\dag} \\ 9^{\dag}
	  \end{array} \right]. \label{Tadd1dag}
          \end{eqnarray}

          The operators $I=T^{\dag}T=\sum_{i=1}^{9}i^{\dag}i$ and
$F=TT^{\dag}=\sum_{i=1}^{9}ii^{\dag}$ where $i$ is the term number are given by
          \begin{eqnarray}
          I & = &  \sum_{j=-\infty}^{\infty}
          [(Q_{0}+Q_{1}+Q_{3})(P_{0,j}+P_{2,j})]+ Q_{2}  \label{Tdom} \\ 
          F & = &  \sum_{j=-\infty}^{\infty}[(Q_{0}(P_{0,j}
          +P_{2,j})+Q_{1}(P_{2,j}+v_{j}P_{0,j}v_{j}^{\dag}))P_{j+1} +
          Q_{2}(P_{0,j}+P_{2,j})P_{j-1}] +Q_{3}. \label{Trng}
          \end{eqnarray}
          Since $I$ and $F$ are projection operators $T$ and $T^{\dag}$ are
          partial isometries with $I$ and $F$ projection operators on the
          domain and range space of $T$.  The role of $I$ and $F$ is
          reversed for $T^{\dag}$ in that $F$ and $I$ are the respective
          domain and range space projection operators for $T^{\dag}$.  The
          $i$ sums over terms show that only the diagonal terms contribute,
          that is $i^{\dag}j =ij^{\dag} =0$ if and only if $i\neq j$.

          The action of $T$ can be shown by consideration of an initial
          lattice qubit state $\vert \underline{s}\rangle$ with qubits at
          sites $0$ and $n+1$ in the state $\vert 2\rangle$ and in the
          state $\vert 0\rangle$ at all other sites.  Start with an initial
          state of the form $\vert 0,-m,\underline{s} \rangle$ with the
          head in state $\vert 0 \rangle$ and at position $-m$.  Iteration
          of $T$ moves the head up to position $0$ by the actions of term
          1. The state of the head and qubit lattice at this point is shown
          in Figure 2.  Term 2 acts once to give the state $\vert
          1,1,\underline{s}\rangle$.  Term 3 takes over  to give the state
          $\vert 1,n+1\rangle \Pi_{j=0}^{n}v_{j}\underline{s}\rangle$ after
          n iterations of $T$.  At this point the lattice qubit state is a
          linear superposition of $2^{n}$ states in the computation basis
          with the coefficients depending on $v$.  If $v=(\sigma_{z}+
          \sigma_{x})/\sqrt{2}$ then the above iteration of $T$ carries out
          the Hadamard transformation on the $n$ qubits.  

          Note that so far, and in the following, at most one term of $T$
          is active at any iteration.  This is shown by Eqs. \ref{Tdom} and
          \ref{Trng} where only the diagonal terms contribute.  In fact $T$
          was constructed so that it has this property as it greatly
          lessens the complication of path determination during iteration
          of $T$ and $T^{\dag}$. 

          The next iteration (term 4) gives the head state $\vert
          2,n\rangle$ with no change in the lattice qubit state.  Terms 5
          and 6, which implement the "add 1" operation, now become active.
          At this point which term is active and the sequence of actions is
          different for the different components in the superposition of
          the $n$ qubit states.  Iteration of term 5 moves the head back
          along a string of $1s$ changing the qubit state $\vert 1\rangle$
          to $\vert 0\rangle$ until a qubit in state $\vert 0\rangle$ is
          encountered.  Term 6 then changes the qubit to state $\vert
          1\rangle$, which completes the "add 1" operation in that
          component state.  On all component states in which the site $n$
          qubit is in state $\vert 0\rangle$, term 5 never acts and term 6
          acts just once.  Note that the site $n$ qubit corresponds to the
          least significant (units) position with significance increasing
          as $n$ decreases to $0$, the most significant position.

          At this point the computation is completed.  Terms 7,8, and 9
          which move the head to the right without stopping and no further
          qubit state changes, now become active.  The final state, when
          "add 1" is just completed in all components, has the following
          form:
          \begin{eqnarray}
          \Psi_{3n+4} & = & \sum_{j=0}^{n}[(1-\delta_{j,n})\langle 0\vert 
 v\vert 0\rangle +\delta_{j,n} ](\langle 1\vert v\vert 0\rangle)^{j} \nonumber 
 \\  & & \mbox{} \times \vert 0,3n+2-
          2j\rangle v\vert 0_{1}\rangle v\vert 0_{2}\rangle \cdots v\vert
          0_{n-1-j}\rangle \vert 1_{n-j}\rangle \vert 0_{n-j+1} \cdots
          \vert 0_{n}\rangle \vert \underline{s}_{\neq [1,n]}\rangle.
          \label{T3n+4}
          \end{eqnarray}
          In this state $\langle i\vert v\vert i^{\prime}\rangle$ is the
          $v$ matrix element between the qubit states $\vert i\rangle
          ,\:\vert i^{\prime}\rangle$ where $i,i^{\prime} =0,1$, and $\vert
          0,3n+2-2j\rangle$ denotes the head in state $\vert 0\rangle$ at
          lattice position $3n+2-2j$. $\vert \underline{s}_{\neq
          [1,n]}\rangle$ is the state of all qubits not at sites
          $1,2,\cdots ,n$. The states of these qubits are unchanged by the
          action of $T$ or $T^{\dag}$.

          The state $\Psi_{3n+4}$ is the state obtained by $3n+4$
          iterations of $T$ on the initial state $\vert 0,0,\underline{s}
          \rangle$ with $\underline{s}$ given by Figure 2.   That is
          $\Psi_{3n+4}=T^{3n+4} \vert 0,0,\underline{s}\rangle$.  All $2^{n}$
          components in the computation basis are included even though
          there are only $n+1$ terms in the j-sum.  The reason is that each
          term in the j-sum corresponds to "add 1" to all $2^{n-1-j}$
          components with $j$ consecutive $1s$ at sites $n,n-1, \cdots , n-
          j+1$ and a $0$ at site $n-j$. (For $j=n$ $2^{n-1-j} =1$.)

          It is of interest to schematically expand the states along the
          computation path as superpositions of states in $B_{C}$.  To this
          end let the state $\Psi_{m}$ denote the path state obtained by m
          iterations of $T$ or $T^{\dag}$ on the state $\vert
          0,0,\underline{s}\rangle$.  That is 
          \begin{equation}
          \Psi_{m}= \left\{ \begin{array}{ll} T^{m}\vert
          0,0,\underline{s}\rangle  & \mbox{if $m\geq 0$} \\(T^{\dag})^{-
          m}\vert 0,0,\underline{s}\rangle &  \mbox{if $m<0$}
          \end{array} \right.  \label{Tm}
          \end{equation}
          The expansion of $\Psi_{m}$  as a superposition of states in
          $B_{C}$ for increasing $m$ shows an initial line branching into
          an n stage binary tree (generated by n iterations of term 3 in
          Eq. \ref{Tadd1}) with each of the $2^{n}$ lines continuing.  This
          is shown in Figure 3 which shows the tree development by
          successive iterations of $T$ starting from the initial state
          shown in Figure 2.  

          The above construction can be extended to initial lattice qubit
          states with more than two qubits in state $\vert 2\rangle$.   For
          example let $\vert \underline{s}\rangle$ be such that qubits at
          sites $0,n_{1},n_{2},n_{3}$ with $0<n_{1}<n_{2}<n_{3}$ are in
          states $\vert 2\rangle$ and are in state $\vert 0\rangle$
          everywhere else.  Let the number of $\vert 0\rangle$ state sites
          between the $\vert 2\rangle$ sites be given by $n =n_{1}-
          1,p=n_{2}-1-n_{1},m=n_{3}-1-n_{2}$.

          In this case the structure of $T$ is such that "add 1" is carried
          out on the $n$ and $m$ qubits in the first and third regions. 
          Iteration of $T$ for states where the head is in the second
          region of $p$ qubits just moves the head along with no changes in
          the head or lattice qubit state  until the head arrives at site
          $n_{2}$.  Note that in Eq. \ref{T3n+4} the head position is
          different in each of the $n$ components in the $j$ sum.  As a
          result the iteration number or arrival time for the head at
          $n_{2}$ is different for each of the $n$ components. 

          The final states with the head to the right of $n_{3}$ describe
          the completed "add 1" in both regions. The structure of the final
          state corresponding to Eq. \ref{T3n+4} 
	  is complex as it contains a double sum $\sum_{j,k}$ with $j
          =0,1,\cdots ,n$ and $k=0,1,\cdots ,m$. Also the head position is
          different for each term in the sum as it depends on both $j$ and
          $k$.  Each component in the double sum is a product of lattice
          qubit states of the form given in Eq. \ref{T3n+4} for $j=j$ and
          for $j=k$.  For $j=k$ the lattice qubit state describes the
          qubits in the region between sites $n_{2}$ and $n_{3}$.
          Adjustments in qubit state indices that label the qubit location
          and for the possibility that $m\neq n$ must be made.  Details are
          left to the reader.

          Expansion of the final states as superpositions of states in
          $B_{C}$ gives a double binary tree structure with a binary tree
          of $m$ stages attached to each terminal branch of an $n$ stage
          tree.  The length of the line (corresponding to the number of $T$
          iterations) between the last branching in the first tree and the
          beginning of branching in the second tree depends on the lattice
          distance $n_{2}-n_{1}$ and to which component in the $j$ sum of
          Eq. \ref{T3n+4} a particular branch corresponds.  The structure
          can be visualized by attaching an $m$ stage binary tree like that
          shown in Figure 3 to each of the $2^{n}$ output paths of the tree
          shown in the figure.  The second tree would begin about  $n_{2}-
          n_{1}$ path states from the solid triangles shown in the output
          paths.  

          For initial lattice qubit states with an even number $2h$ of
          qubits in state $\vert 2\rangle$ with regions of $0s$ in between
          the above attachment of binary trees to the end branches of a
          preceeding tree is iterated $h$ times.  The lengths of the lines
          betweeen the last branching of the $kth$ tree and the beginning
          of the $k+1st$ tree for $k=1,2,\cdots h-1$ follow a pattern
          similar to that described above for the two trees $h=1$.

          The terminal tree structure is different for initial qubit states
          that contain an odd number of qubits in state $\vert 2\rangle$. 
          to see this it is sufficient to consider the case with the site
          $0$ qubit the only one in state $\vert 2\rangle$ and all other
          qubits in state $\vert 0\rangle$.  In this case only the first
          three terms of $T$ are active.  Iteration of $T$ moves the head
          in state $\vert 1\rangle$ to the right without stopping. As it
          moves it converts each qubit from state $\vert 0\rangle$ to
          $v\vert 0\rangle$. 

          Expansion of the path states in terms of the basis $B_{C}$  shows
          a nonterminating binary tree.  The branching begins when term 3
          of Eq. \ref{Tadd1} becomes and remains permanently active.  For
          initial states with an odd number $2h+1$ of qubits in state
          $\vert 2\rangle$ separated by regions of $0s$.  There are $h$
          iterations of finite stage binary trees attached as described
          above for the even number of $2s$ in the initial state.  In
          addition a nonterminating binary tree is attached to each
          terminal branch of the last  ($hth$) tree in the iteration.

          It remains to discuss the proof that $T$ and $T^{\dag}$ are
          distinct path generating.  This is done by considering the path
          states $\Psi_{m}$ defined above in Eq. \ref{Tm}. One has to show
          that for each $m\geq 0$ iteration of $T^{\dag}$ on $\Psi_{m}$
          does not generate new states.  That is for each $n<0$
          $(T^{\dag})^{-n} \Psi_{m} =\Psi_{m+n}$.  Also for each $m<0$ and
          each $n>0$ $T^{n}\Psi_{m}=\Psi_{m+n}$.   The critical step in the
          proof is to note that $T^{\dag}$, acting on the state $\langle
          0\vert v\vert 0\rangle \vert 3,j+1\rangle \vert 1_{j}\rangle
          +\langle 1\vert v\vert 0\rangle \vert 2,j-1\rangle \vert
          0_{j}\rangle$, gives (terms $5^{\dag},\: 6^{\dag}$ active) the
          state  $\vert 2,j\rangle  v\vert 0_{j}\rangle$.  Irrelevant parts
          of the state have been excluded for clarity. Additional details
          are left to the reader.

          It is clear that this proof applies to initial states containing
          arbitrary distributions of $0s$ and $2s$ in the qubit state
          lattice and to arbitrary positions of the head in any one of the
          four internal states.  For all of these states $T$ is a direct
          sum of countable many copies of the bilateral shift and the
          restriction of $T$ to these subspaces is unitary.  

          The proof is easily extended to cover initial states with one or
          more qubits in state $\vert 1\rangle$. Thus it is clear that $T$
          is distinct path generating in that it is a direct sum of  copies
          of finite shifts, unilateral shifts, complements of unilateral
          shifts, and bilateral shifts.  

          \subsection{QTMS with Interferometer Graph Structures} 

          The above example shows that the restriction of $T$ to be
          distinct path generating is sufficiently powerful to include
          quantum computations that generate branchings in $B_{C}$ where
          the specific computation steps in each branch are different.  The
          graph structure for a large number in intial lattice qubit states
          was seen to be that of iterated binary trees with or without a
          nonterminating terminal tree.

          Here two examples of QTMs are given for which some paths in
          $B_{T}$ correspond  to graphs in $B_{C}$ that have the structure
          of interferometers.  It is of interest to note that use of
          coordinate space interferometers to factor integers has been
          described \cite{ClDo}.  

          For the first example let $T$ be given as an $l,s$ sum, Eq.
          \ref{Tls}, by
          \begin{equation}
          T=\sum_{j=-
          \infty}^{\infty}[Q_{0}P_{0,j}+wQ_{0}P_{1,j}+(Q_{1}+Q_{2})P_{0,j}+
          Q_{0}w^{\dag}P_{1,j}]uP_{j} \label{Tintrfr1}
          \end{equation}
          Here $w$ is a unitary operator that takes the head state $\vert
          0\rangle$ into the linear superposition $(1/\sqrt{2})(\vert
          1\rangle+\vert 2\rangle)$ of head states $\vert 1\rangle$ and
          $\vert 2\rangle$.  It is this linear superposition that creates
          the two branches of the interferometer.  The interferometer is
          opened when the second term of $T$ is active and closed when the
          last (righthand) term is active.

          It is straightforward to prove that $T$ is a power partial
          isometry with all types of noncyclic paths present in the
          decomposition into shifts. Note that any state of the form $\vert
          \phi ,j,\underline{s}\rangle$ where $\phi$ is the head state
          $(1/\sqrt{2})(\vert 1\rangle -\vert 2\rangle$ and the site $j$
          qubit is in state $\vert 1\rangle$ is a terminal $B_{T}$ path
          state for $T$.

          Iteration of $T$ on any state of the form $\vert
          0,j,\underline{s} \rangle$ where $j<1$ and $\vert \underline{s}
          \rangle$ has the form shown at the top of Figure 4, moves the
          head to the right until the state $\vert 1\rangle$ qubit is
          encountered at site $1$.  The next iteration opens the
          interferometer with the two branches labelled by different head
          states.  Continued iteration of $T$ describes head motion along
          the lattice with the head in state $(1/\sqrt{2})(\vert 1\rangle
          +\vert 2\rangle)$.  This state, which is a coherent sum of 2
          $B_{C}$ states, describes the two arms of the interferometer. 
          The interferometer is closed when the second qubit in state
          $\vert 1\rangle$ at site $n$ is encountered.  The bottom of
          Figure 4 shows the interferometer where paths of connected dots
          refer to paths of states in $B_{C}$.  The branches labelled with
          $2$ and $1$ correspond to the two head states $\vert 2\rangle$
          and $\vert 1\rangle$.  Note that the positions of opening and
          closing of the interferometer correspond to the lattice positions
          of the qubit in state $\vert 1\rangle$.  This occurs because all
          terms of $T$ show the head moving in the same direction. 

          This example is a very simple QTM in which the action in both
          branches of the interferometer is identical and no changes in the
          qubit lattice occur.  The two arms are distinguished by different
          head states only.  It is of interest to see if there exist QTMs
          for which different activities occur in the two arms and for
          which the step operators are distinct path generating.  The
          answer is that such QTMs exist provided activities in each arm
          are such that all differences in the computation states in the
          two arms generated by iteration of $T$ are removed at the step
          prior to coherent combination or closure of the arms.  Also this
          must hold for all initial states.

          An example of such a QTM is given by the following operator with 9
terms:
          \begin{eqnarray}
          T & = & \sum_{j=-\infty}^{\infty}\left[ \begin{array}{c} 
          Q_{0}P_{0,j}u \\ 1 \end{array} +\begin{array}{c} 
          w^{\prime}_{12}Q_{0}P_{1,j}u \\ 2 \end{array} +\begin{array}{c} 
          w^{2}Q_{1}vP_{0,j}u^{\dag} \\ 3 \end{array} + \begin{array}{c} 
	  w^{2}Q_{2}P_{0,j}u^{\dag} \\ 4 \end{array}
   + \begin{array}{c} w^{2}Q_{3}P_{1,j}u \\ 5 \end{array} \right. \nonumber \\
  & & \mbox{} \left. + \begin{array}{c} w^{2}Q_{4}P_{1,j}u \\ 6 \end{array} + 
  \begin{array}{c} w^{2}Q_{5}P_{0,j}v^{\dag}u \\ 7 \end{array} + 
  \begin{array}{c} w^{2}Q_{6}P_{0,j}u \\ 8 \end{array} +\begin{array}{c}
          Q_{0}(w^{\prime}_{78})^{\dag}P_{1,j}u \\ 9 \end{array} \right] P_{j}
          \label{Tintf2}
          \end{eqnarray}
          As before the term numbers are placed below the terms.  Terms
3, 5, and 7 are active in one arm  and terms 4, 6, and 8 are active in the
other arm of the interferometer.  Terms 2 and 9 open and close the
interferometer.

          There are 9 different head states with $w$ the unitary shift
          shift operator on the head states defined by $wQ_{h}=Q_{h+1}w
          \bmod 9$.  The two unitary operators $w^{\prime}_{12},\:
          w^{\prime}_{78}$ satisfy $w^{\prime}_{12}\vert 0\rangle =
          1/\sqrt{2}(\vert 1\rangle +\vert 2\rangle )$ and 
          $w^{\prime}_{78}\vert 0\rangle = 1/\sqrt{7}(\vert 1\rangle +\vert
          8\rangle )$.  The two dimensional unitary operator $v$ is arbitrary,
          and the head shift operator $u$ is as defined in other examples.

          The interferometer is opened by the action of term 2 where the
          head in state $\vert 0\rangle$ is converted to state
          $1/\sqrt{2}(\vert 1\rangle +\vert 2\rangle )$.  Terms 3,5,and 7
          are active in succession in the arm  with head
          state $\vert 1\rangle$, and terms 4,6, and 8 are active in
          succession in the arm  with head state $\vert
          2\rangle$.  The interferometer is closed by the action of term 9. 
          Each arm of the interferometer is a path containing just 4
          states.   This is a consequence of the fact that, except for the
          first term, each term of $T$ in an iteration is active just once.
          This follows from the presence and properties of the head state
          change operators present in each term except the first.

          The activities in the two arms are different because of the
          presence of $v$ and $v^{\dag}$ in terms 3 and 7 but not in 4 and
          6.  Thus the middle two states in the arms differ by both the head
          states and the qubit states on which terms 3 and 5, and 4 and 6
          act.  These differences are shown in Figure 5 which shows
          explicitly the computation basis states in the interferometer
          arms for an initial state given by $\vert 0,0\rangle \vert
          1_{1},0_{2},1_{3}\rangle \vert \underline{0}_{else}\rangle$. 
          This state has the head in state $\vert 0\rangle$ at lattice site
          $0$ and the qubit lattice state with $0s$ everywhere except at
          sites $1$ and $3$. The states of the qubits at sites $1,2,3$ are
          shown explicitly.    

          The production of interferometers described above can be iterated
          by suitable choice of initial lattice qubit states.  For example
          a string of $n$ interferometers is generated for the initial
          state $\vert 0,0,\underline{s}\rangle$ where $\vert
          \underline{s}\rangle$ contains n repetitions of the pattern
          $\vert 1_{j},0_{j+1},1_{j+2}\rangle$ for $j>0$ with each
          repetition separated by one or more $0s$.  

          It is straight forward to prove that $T$ is distinct path
          generating with all types of shifts present.  One way to do this
          is to show that iteration of $T^{\dag}$ on $T^{n}\psi$ for
          various $n$ and different initial states does not generate new
          states (i.e. $(T^{\dag})^{m}T^{n}\psi =T^{n-m}\psi$ for $m\leq
          n$.  Because of the structure of the terms of $T$  it is
          sufficient to show this for values of $n =0,1,\cdots 7$ for
          suitably chosen states.  The same demonstration is also needed
          for iterations in which the roles of $T$ and $T^{\dag}$ are
          interchanged.

          As noted for this example the head motion is exactly the same for
          both arms of the interferometer.  That is terms 3 and 4 move the
          head one site to the right. Terms 5,6,7, and 8 all move the head
          one site to the left.  It is of interest to see what happens if
          this synchrony of head motion in the arms is broken.  An example
          would be to change the definition of $T$ by exchanging $u$ and
          $u^{\dag}$ in terms 4 and 6.  In this case it is easy to show
          that the altered $T$ is not distinct path generating.  The reason
          is that under the alteration term 6 scans a different qubit in
          state $\vert 1\rangle$ than does term 4.  It is then easy to see
          that iteration of $T$ on initial states of the form $\vert
          0,0\rangle \vert 1_{j}\rangle \vert \underline{0}_{\neq
          j}\rangle$ for $j>0$ generates terminal states in both arms of
          the interferometer.  However the path in the arm in which the
          altered term 6 is active is shorter than the path in the arm in
          which term 5 is active. Iteration of $T^{\dag}$ on any state
          representing just one arm of the interferometer will not
          regenerate the initial state.

          This example illustrates the fact that there are step operators
          $T$ that are distinct path generating on some subspace of $\cal
          H$ even if they are not distinct path generating on the whole
          space.  The above example showed states on which $T$ is not
          distinct path generating.  However this altered $T$ is distinct
          path generating on lattice qubit states of the form $\vert
          1_{j},0_{j+1},1_{j+2}\rangle$ and $0$ everywhere else with the
          head in state $\vert 0,h\rangle$ with $h<j$. Besides these states
          there are many other types of initial states on which $T$ is
          distinct path generating.

          \section*{Acknowledgements}
          This work is supported by the U.S. Department of Energy, Nuclear 
          Physics Division, under contract W-31-109-ENG-38.

          \newpage

          \begin{center}
          FIGURE CAPTIONS
          \end{center}

          Figure 1.  Different types of Possible Paths in a Basis $B_{T}$
          for which $T$ is Distinct Path Generating.  The small dots denote
          path states and the large dots path states that are terminal. 
          Infinite paths are labelled by a, b, and c and finite paths by d
          and e.  The corresponding shift types are a, bilateral (unitary);
          b, unilateral (isometry); c, adjoint of unilateral shift
          (coisometry); d, finite shift (noncyclic); and e, cyclic shift
          (unitary). \\
          \\
          Figure 2. An Initial State for the "Add 1" Quantum Turing
          Machine.  The vertical arrow with a $0$ at site $0$ denotes the
          head in state $\vert 0\rangle$ at site $0$.  The two qubits in
          state $\vert 2\rangle$ at sites $0$ and $n+1$ are shown, with all
          other qubits in state $\vert 0\rangle$. \\
          \\
          Figure 3.  The n Stage Binary Tree of Paths of States in $B_{C}$
          for the "Add 1" Quantum Turing Machine. Each  dot denotes a state
          in $B_{C}$.  The vertical arrow at the tree root denotes the
          computation basis state shown in Figure 2.  The numbers at each
          vertex give the state of the qubit at the head location at each
          stage. The solid arrowheads denote the path state at which the
          "add 1" operation is complete and term 1 of $T$, Eq. \ref{Tadd1}
          becomes active.  The location of the arrowhead in each path
          depends on the number $1s$ before a $0$ is reached.  No arrowhead
          is shown for the uppermost path (all $1s$) as it occurs off the
          right edge of the figure at the path state $3n+4$ steps from the
          tree root state. \\
          \\
          Figure 4.  Initial Qubit Lattice State and Paths in $B_{C}$ for
          the Interferometer Quantum Turing Machine.  The initial latttice
          stae in the upper part of the figure shows qubits in state $\vert
          1\rangle$ at sites $1$ and $n$ and in state $\vert 0\rangle$
          elsewhwere. The lower part of the figure shows the
          interferometric structure of the $B_{C}$ paths. The numbers $1,2$
          denote the two different head states that distinguish the two
          arms of the interferometer.  The interferometer opens when the
          head is at site $1$ and closes when the head is at site $n$.  The
          number of states in the interferometer arms equals $1$ plus the
          number of state $\vert 0\rangle$ qubits between the $1s$ because
          all terms in $T$ move the head in the same direction (from left
          to right). \\
          \\
          Figure 5. Path States in $B_{C}$ for the Quantum Turing Machine
          Interferometer with $T$ given by Eq. \ref{Tintf2}. The states are
          shown explicitly.  Normalization coefficients ($1/\sqrt{2}$) are
          excluded. For the middle two states the state of the site $2$
          qubit is different in the two arms.

          
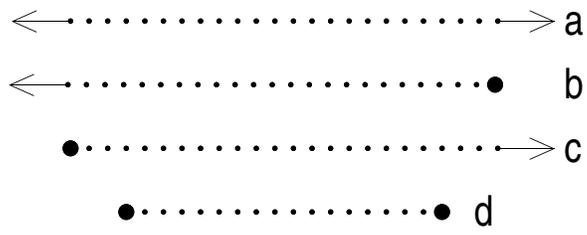
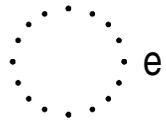
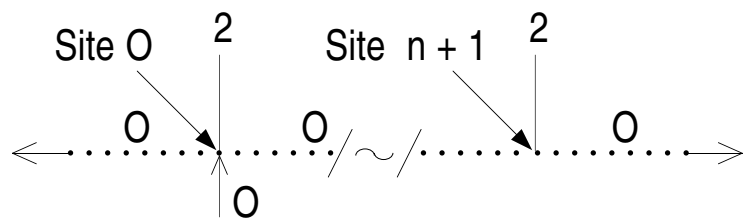
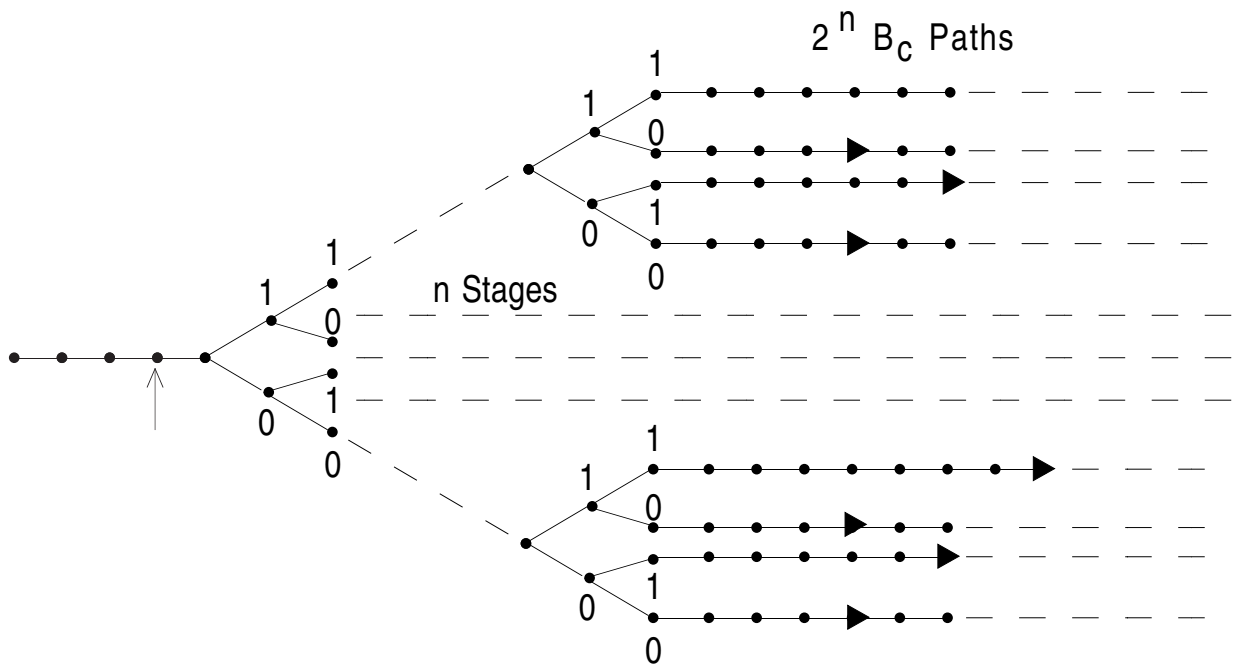
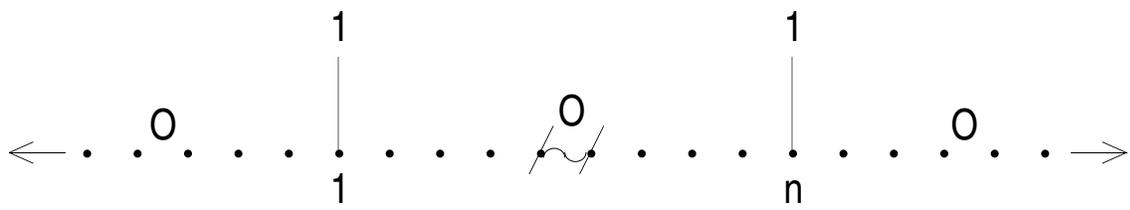
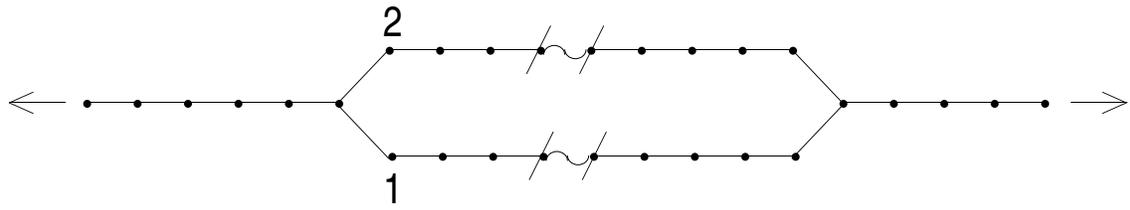
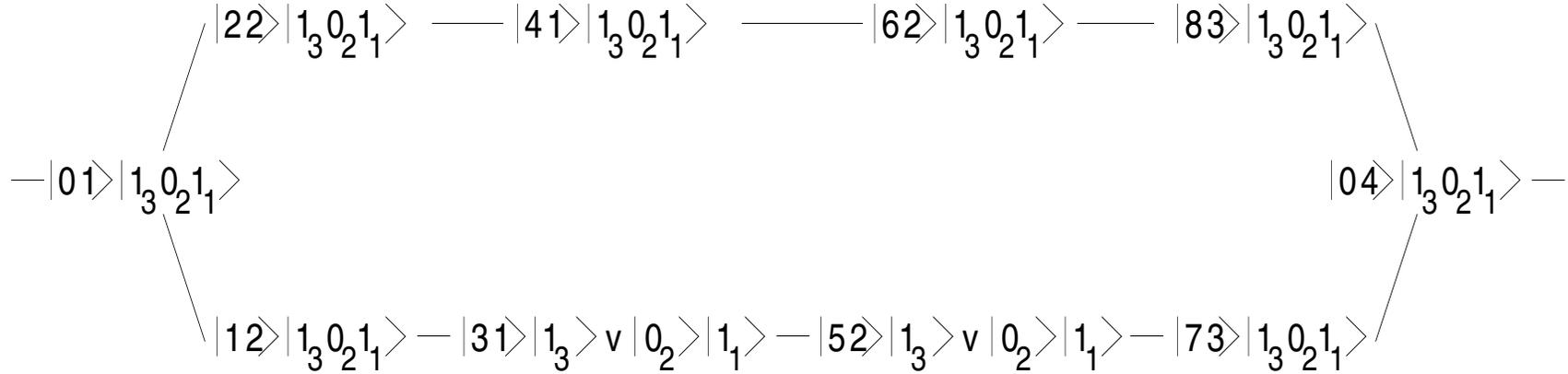
\end{document}